\documentclass[twocolumn,showpacs,preprintnumbers,amsmath,amssymb]{revtex4}

\usepackage{graphicx}
\usepackage{dcolumn}
\usepackage{bm}
\newcommand{\gtsim}{\mbox{{\raisebox{-0.4ex}{$\stackrel{>}{{\scriptstyle\sim
}}
$}}}}
\newcommand{\ltsim}{\mbox{{\raisebox{-0.4ex}{$\stackrel{<}{{\scriptstyle\sim
}}
$}}}}
\def\cuscn{$\kappa$-(BEDT-TTF)$_2$Cu(NCS)$_2$}
\begin{document}
\title{Persistence to high temperatures of
interlayer coherence in an organic superconductor}
\author{John Singleton$^1$, PA~Goddard$^{1,2}$,
A~Ardavan$^2$, AI~Coldea$^{3}$, SJ~Blundell$^2$,
RD~McDonald$^1$,
S~Tozer$^4$,
JA~Schlueter$^5$}
\affiliation{$^1$National High Magnetic Field Laboratory,
MPA-NHMFL, TA-35, MS-E536, Los Alamos National Laboratory,
Los Alamos, NM87545, USA\\
$^2$Department of Physics,
University of Oxford, The Clarendon Laboratory,
Parks Road, Oxford~OX1~3PU, UK\\
$^3$H.H.~Wills Physics Laboratory, Bristol University,
Tyndall Avenue, Bristol, BS9 1TL, UK\\
$^4$NHMFL, 1800 E. Paul Dirac Drive, Tallahassee FL~32310, USA\\
$^5$Materials Science Division, Argonne National Laboratory,
Argonne, Illinois 60439, USA
}
\date{\today}
\begin{abstract}
The interlayer magnetoresistance $\rho_{zz}$
of the organic
metal \cuscn ~is studied in
fields of up to 45~T and at temperatures $T$
from 0.5~K to 30~K.
The peak in $\rho_{zz}$
seen in in-plane fields, a definitive
signature of interlayer
coherence, remains to $T$s exceeding the Anderson
criterion for incoherent transport by a
factor $\sim 30$.
Angle-dependent
magnetoresistance oscillations
are modeled using an approach based on
field-induced quasiparticle paths on
a 3D Fermi surface, to yield the
$T$ dependence of the
scattering rate $\tau^{-1}$.
The results suggest that
$\tau^{-1}$ does not vary strongly over the Fermi surface,
and that it has a $T^2$ dependence due to
electron-electron scattering.
\end{abstract}


\maketitle
\begin{figure}[htbp]
\vspace{-5mm}
\centering
\includegraphics[width=8.5cm]{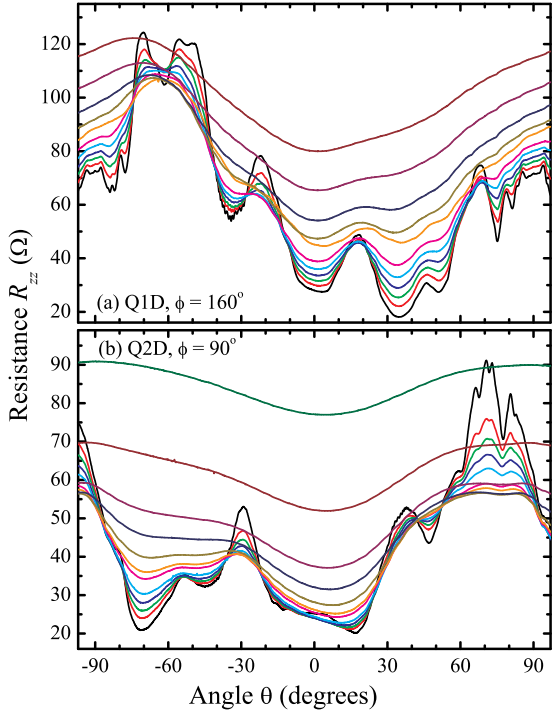}
\vspace{-5mm}
\caption{Interlayer resistance $R_{zz}(\propto \rho_{zz})$ of a
\cuscn ~crystal versus tilt angle $\theta$ for various
constant $T$; $B=45$~T. (a)~Data for $\phi = 160^{\circ}$, a plane
of rotation at which $\rho_{zz}$ is determined by phenomena
on the Q1D FS sections. In order of
increasing $R_{zz}$ at $\theta = 35^{\circ}$, the curves are for $T
= 5.3$, 6.5, 7.6, 8.6, 9.6, 10.6, 12.1, 13.1, 14.6, 17.1, 19.6 and
29.3~K respectively. In addition to AMROs,
DKCOs can be seen as twin peaks on either
side of $\theta=90^{\circ}$ in the lower $T$ data.
(b)~Similar data for $\phi=90^{\circ}$; here
$\rho_{zz}$ features are associated with the Q2D FS
section. In order of increasing $R_{zz}$ at $\theta = -70^{\circ}$,
the curves are for $T=5.3$, 7.6, 8.6, 9.6, 10.6, 12.1, 13.1, 14.6,
17.1, 19.6, 24.5 and 29.3~K respectively.}
\label{fig1}
\vspace{-5mm}
\end{figure}
\begin{figure}[htbp]
\vspace{-5mm}
\centering
\includegraphics[width=9cm]{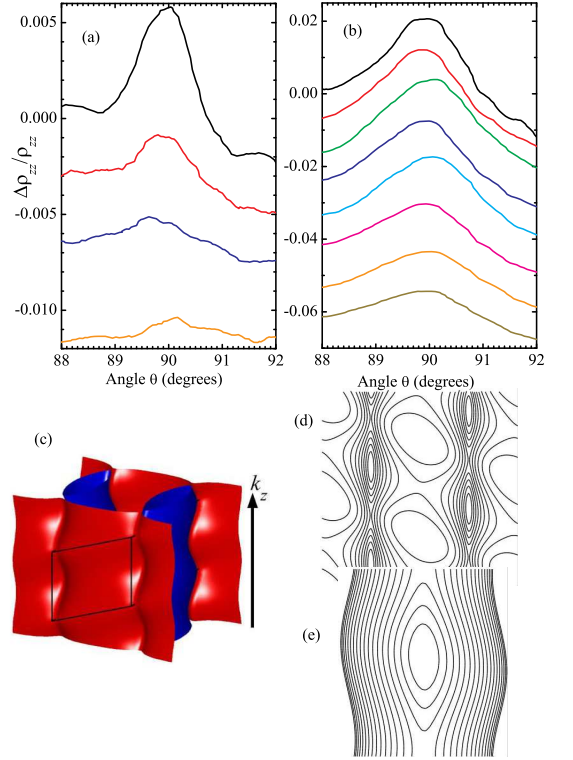}
\vspace{-5mm}
\caption{(a)~The 45~T magnetoresistance close to 
$\theta=90^{\circ}$ at
$\phi=160^{\circ}$, plotted as 
$\Delta \rho_{zz}/\rho_{zz{\rm BG}}$, the
fractional change in $\rho_{zz}$ from the more slowly-varying
background. Data for $T=5.3$ (highest), 7.6, 9.6 and 13.1~K (lowest)
are shown, offset for visibility. (b)~Similar data for
$\phi=90^{\circ}$; the traces are for $T=5.3$ (highest), 
7.6, 8.6, 9.6, 10.6, 12.1, 13.1 
and 14.6~K (lowest). Each trace has been
offset for clarity. (c)~3D representation of the FS of
\cuscn (after Ref.~\cite{goddardprb2004}); the finite
$t_{\bf a}$ gives the corrugations (shown greatly
exaggerated) on the sides of the FS. 
Q1D and Q2D FS sections are shown in red and blue
respectively. (d)~Consequent
field-induced closed orbits on the side of the Q1D FS
section when $\theta=90^{\circ}$ and $\phi=0$. (e)~Similar closed
orbits on the Q2D FS section when
$\theta=90^{\circ}$ and $\phi=90^{\circ}$. Orbits such as those in
(d) and (e) give rise to the SQUIT peak in $\rho_{zz}$.}
\label{fig2}
\vspace{-5mm}
\end{figure}

The past two decades have seen a tremendous blossoming
of interest in compounds that
possess quasi-two-dimensional (Q2D)
electronic bandstructure;
examples include crystalline organic
metals~\cite{mckenzie,osada,lebed,singleton},
cuprates~\cite{hussey}
and layered ruthenates~\cite{eva}.
These materials may be described by a tight-binding Hamiltonian
in which the ratio of the interlayer transfer integral $t_{\perp}$
to the average intralayer transfer integral
$t_{||}$ is $\ll 1$~\cite{mckenzie,osada,lebed,singleton}.
The question arises as to whether the interlayer charge
transfer is coherent or incoherent in these materials,
{\it i.e.} whether or not
the Fermi surface (FS) is three dimensional
(3D), extending in the interlayer
direction.
Various criteria for interlayer incoherence
have been proposed, including~\cite{anderson}
\begin{equation}
k_{\rm B}T > t_{\perp},
\label{philanderson}
\end{equation}
where $T$ is the temperature. In this picture, thermal
fluctuations ``wipe out'' details of the interlayer
periodicity~\cite{anderson}. The consequent interlayer 
incoherence is used as a
justification for a number of theories which are thought to be
pivotal in the understanding of Q2D materials
(see e.g. \cite{anderson,strong,review,bristoltheory}). It is
therefore important to test assertions such as
Eq.~\ref{philanderson}. To this end, we have measured
the magnetic-field-orientation dependence of the resistance of the
organic metal
$\kappa$-(BEDT-TTF)$_2$Cu(NCS)$_2$~\cite{review} using fields of up
to 45~T and $T$s of up to 30~K (Fig.~\ref{fig1}). This
material was chosen because its FS is well
known~\cite{review,goddardprb2004} (see Fig.~\ref{fig2}(c)), 
previous low-$T$
experiments have shown that the interlayer transfer integral
is $t_{\bf a}=0.065 \pm 0.007$~meV~\cite{goddardprb2004,footnote}
and modest $T$s allow the inequality in
Eq.~\ref{philanderson} to be exceeded by orders of magnitude
($t_{\bf a}/k_{\rm B}\approx 0.5$~K). Our data show that interlayer
coherence survives to at least $k_{\rm B}T~\gtsim 30t_{\bf a}$ 
({\it c.f.} Eq.~\ref{philanderson}). In addition, the results in
this paper determine the low-$T$ scattering
rate $\tau^{-1}$ for the quasiparticles. Recent works stress that an
understanding of the effects of the scattering mechanism in organic
metals gives vital information about the
mechanism for superconductivity~\cite{mckenzie2005,analytis}.

\begin{figure}[htbp]
\vspace{-5mm}
\centering
\includegraphics[width=8.5cm]{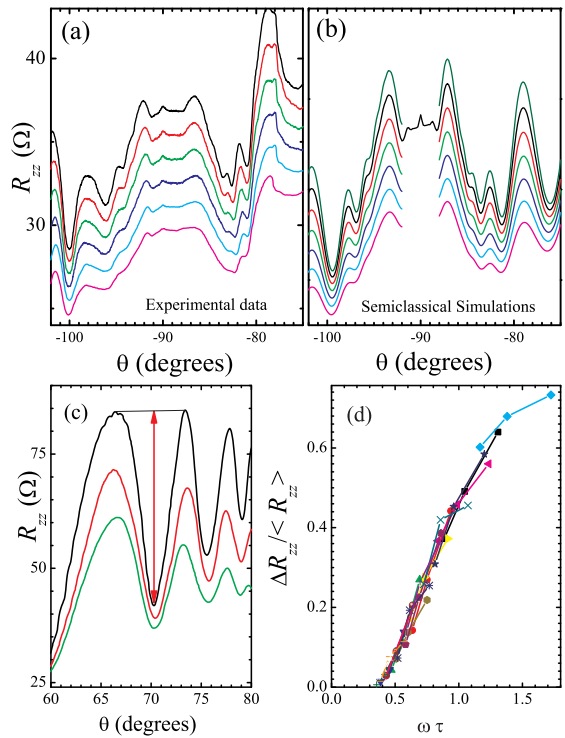}
\vspace{-5mm}
\caption{Comparison of experimental $\rho_{zz}$ data~(a) and 
the numerical simulation~(b) method described in
Ref.~\cite{goddardprb2004}. $T=1.5$~K and $\phi = 15^{\circ}$; for
the lowest to the highest traces, $B$ is 32, 34, 36, 38,
40, 42 and 44~T (no offset is applied). (c)~AMRO data for
$\phi=150^{\circ}$, $T=1.5$~K and fields of 20 (lowest), 24 and 28~T
(highest). The arrow indicates the amplitude $\Delta R_{zz}$
of a particular AMRO
feature; $<R_{zz}>$ is the resistance at the arrow's midpoint.
(d)~Experimental AMRO $\Delta R_{zz}/<R_{zz}>$ 
for $\phi=150^{\circ}$ plotted as a
function of the orbit angular frequency $\omega$ times scattering
time $\tau$. Data for several AMROs, denoted by their Yamaji
indices~\cite{review} $i$ are shown ($i=2$ (square), $i=3$ (dot),
$i=4$ (triangle) and $i=5$ (inverted triangle)) for several
$T$s in the range 1.7~K to 5.5~K.}
\label{fig3}
\vspace{-5mm}
\end{figure}

Electrical contacts were made to single crystals ($\sim 0.7\times
0.5\times 0.1$~mm$^3$) of \cuscn
~using $12.5~\mu$m Pt wire bonded by graphite paste; contact
resistances were $\sim 10~\Omega$. Current and voltage terminals
were arranged such that the measured resistance $R_{zz}$ is
proportional to the interlayer component of the
resistivity tensor, $\rho_{zz}$~\cite{goddardprb2004}. 
The measurements
in Figs.~\ref{fig1} and \ref{fig2} use a
well-characterized crystal also 
employed in Refs.~\cite{singleton,goddardprb2004,condprob}. 
Additional experiments
(e.g. Fig.~\ref{fig3}) used d8-\cuscn ~crystals;
here, ``d8'' indicates that the terminal hydrogens of BEDT-TTF
have been replaced by deuterium. These crystals were
studied in Ref.~\cite{goddardprb2004}, where it was
found that deuteration reduces $t_{\bf a}$
to $0.045\pm 0.005$~meV. Samples were
mounted on a ceramic holder attached to a cryogenic goniometer
allowing continuous rotation in 
static magnetic fields {\bf B}. Crystal
orientations are labeled by the angles $\theta, \phi$. Here,
$\theta$ is the angle between {\bf B} and the normal to the
highly-conducting planes and $\phi$ defines the plane of rotation;
$\phi = 0$ is a plane of rotation containing the $k_{\bf b}$
direction and the normal to the highly-conducting
planes~\cite{goddardprb2004}. Sample $T$s were stabilized
using a calibrated Cernox sensor mounted on the
sample holder and a heater driven by a feedback circuit. The Cernox
resistance was monitored during isothermal field sweeps in
order to deduce the correction to the apparent $T$ due to
magnetoresistance; results were in good agreement
with the study of Ref.~\cite{brandt}. Static
fields were provided by the 45~T Hybrid 
magnet at NHMFL Tallahassee
and 33~T Bitter magnets at Tallahassee and HFML Nijmegen.

Typical $R_{zz}$ data are shown in 
Figs.~\ref{fig1}(a) ($\phi
=160^{\circ}$) and \ref{fig1}(b) 
($\phi = 90^{\circ}$) for $T$s in
the range 5.3~K to 29.3~K. 
In the former rotation plane, features in
$\rho_{zz}$ are known to be due to 
orbits on the Q1D FS
sections~\cite{goddardprb2004}, 
whereas in the latter they can be confidently
attributed to Q2D FS
orbits~\cite{goddardprb2004}. A series of
angle-dependent magnetoresistance
oscillations (AMROs), periodic in 
$\tan\theta$~\cite{review,goddardprb2004} is observed 
at both $\phi$; those due to
the Q2D FS sections are frequently known as
{\it Yamaji oscillations}~\cite{review}.
In addition, {\it Danner-Kang Chaikin 
oscillations} (DKCOs)~\cite{goddardprb2004,dkc} are observed
as peaks either side of 
$\theta=\pm90^{\circ}$ when $\phi = 160^{\circ}$. As $T$
increases, the AMROs decrease in intensity, with the higher-index
oscillations ({\it i.e.} those closer to
$\theta=90^{\circ}$~\cite{goddardprb2004}) disappearing first.
After most AMROs disappear, the slowly-varying
background magnetoresistance begins to increase more markedly.

Figures~\ref{fig2}(a) and (b) show expansions of data in
Fig.~\ref{fig1} close to $\theta=90^{\circ}$ at selected $T$s. 
For ease of comparison, data are
plotted as $\Delta \rho_{zz}/\rho_{zz{\rm BG}}$, 
the fractional deviation in
$\rho_{zz}$ from the background magnetoresistance 
$\rho_{zz{\rm BG}}$
determined by fitting a smoothly-varying curve through the data on
either side of $\theta=90^{\circ}$~\cite{singleton}. In such
measurements, interlayer coherence is detected using a phenomenon
known as the ``coherence peak'' or ``SQUIT (Suppression of
QUasiparticle Interlayer
Transport) peak''~\cite{mckenzie,osada,singleton,review,goddardprb2004}, 
a maximum in $\Delta \rho_{zz}/\rho_{zz{\rm BG}}$ 
observed when {\bf B} lies exactly in the
intralayer plane ({\it i.e.} at $\theta=90^{\circ}$). This occurs
because of the efficient interlayer velocity averaging caused by
closed orbits on the side of the FS
(Fig~\ref{fig2}(c-e)); these can exist 
if, {\it and only if}~\cite{mckenzie,singleton}, the
interlayer transport is coherent, {\it i.e.} the FS is a
3D entity extending in the interlayer direction. In both
Figs.~\ref{fig2}(a) and (b), the SQUIT peak is visible close
to $\theta =90^{\circ}$. In spite of the small size of $t_{\bf a}$,
this demonstration of interlayer coherence continues to be
observable up to at least 14.6~K, exceeding the criterion in
Eq.~\ref{philanderson} by a factor $\sim 30$~\cite{vestige}.

The AMROs and DKCOs
in Fig.~\ref{fig1} are very similar to
those observed at $T=0.5$~K~\cite{goddardprb2004}, 
the only difference being that they
decrease in amplitude as $T$ increases;
{\it i.e.} the magnetoresistance features do not change in 
form or angular position
as $T$ grows, they merely fade gradually. This suggests that
the same mechanism is responsible 
for the form of the magnetoresistance
at all $T$
examined. Moreover, the SQUIT peak (Fig.~\ref{fig2})
demonstrates that the FS of \cuscn ~remains 3D up to at
least $T\approx 15$~K~\cite{vestige}. We therefore choose to simulate
the data using the numerical method based on
field-induced quasiparticle trajectories across a 3D FS
that was used to model $T=0.5$~K AMRO results successfully in
Ref.~\cite{goddardprb2004}. This procedure employs Chambers'
Equation and the experimentally-deduced FS
(Fig.~\ref{fig2}(c)~\cite{goddardprb2004}); given {\bf B}
and the angles $\theta,\phi$, the only input
parameter required to calculate $\rho_{zz}$ is
$\tau$. The model has the
restriction that magnetic breakdown 
between the Q1D and Q2D FS sections~\cite{neilbd} is not
included; recently it was shown that breakdown
in \cuscn ~can produce a series
of AMROs~\cite{bangura}. However, the
breakdown amplitude falls off rapidly with increasing
$\theta$~\cite{neilbd}, so that these effects are 
small for $B\approx 45$~T
when $\theta ~\gtsim ~70^{\circ}$~\cite{bangura}. In the
following, we neglect data at lower $\theta$ for
this reason.

A typical comparison of experiment and simulation is shown in
Figs.~\ref{fig3}(a) and (b) ($\phi =15^{\circ}$). For each
$\phi$, it is possible to map the normalized AMRO
amplitude $\Delta R_{zz}/<R_{zz}>$ (see Fig.~\ref{fig3}(c)) from the
simulation onto the experimental value by adjusting 
$\tau$. Moreover, for each $T$, it was found that
the same $\tau$, to within experimental errors, fitted all the
AMRO features.
Given this result, it is useful to find a single
parameter to describe {\it all} AMRO amplitudes for a particular
$\phi$. A good candidate is $\omega \tau$, where
$\omega$ is the angular
frequency at which the orbit responsible for the AMRO feature is
traversed. In the case of
AMROs due to the Q2D FS sections, 
$\omega = (eB/m^*_{\alpha})\cos\theta$, the cyclotron 
frequency~\cite{goddardprb2004}.
Here, $m^*_{\alpha}$ is the $\theta=0$ effective mass of the
$\alpha$ Q2D pocket of the Fermi
surface~\cite{review,neilbd}. 
For orbits on the Q1D Fermi
surface, the relevant period is the time to cross the 
Brillouin zone~\cite{ardavan}. 
As long as one avoids the region close to
$\theta=90^{\circ}$ where the SQUIT and DKCOs
occur (the latter involving orbits that ``snake'' 
across contours on
the FS~\cite{goddardprb2004,dkc}), 
this frequency
can be expressed as
$\omega=(2eB/(m^*_{\beta}-m^*_{\alpha}))\cos\theta$, where
$m^*_{\beta}$ is the $\theta=0$ effective mass 
of the ``$\beta$''
breakdown orbit~\cite{neilbd}.

\begin{figure}
\vspace{-5mm}
\centering
\includegraphics[width=8.0cm]{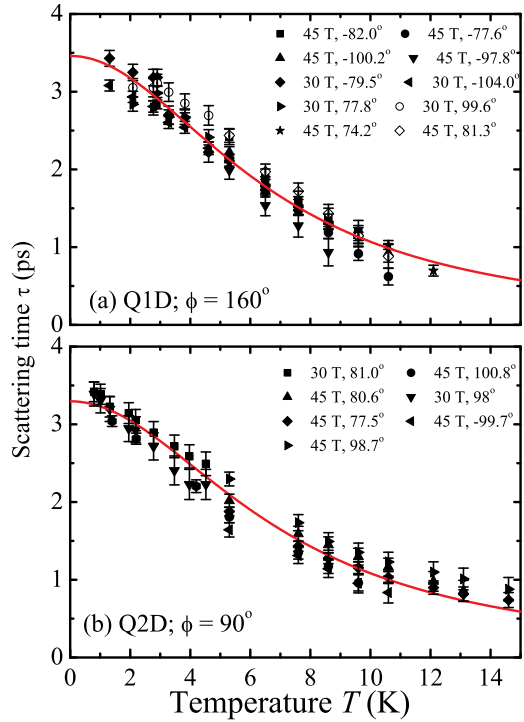}
\vspace{-5mm}
\caption{Scattering time $\tau$
deduced from AMROs at
$\phi \approx 160^{\circ}$~(a)
and $\phi \approx 90^{\circ}$~(b)
versus $T$. Consistent values of $\tau$
are deduced from several
AMRO features at
fields of 30~T and 45~T (insets
show field and $\theta$ for each feature).
The data are fitted to $\tau = (\tau_0^{-1}+AT^2)^{-1}$
(curves) with $\tau_0=3.3 \pm 0.1$~ps
and $A=0.0062 \pm 0.0003~{\rm ps^{-1}K^{-2}}$~(a)
and $\tau_0=3.5 \pm 0.1$~ps
and $A=0.0065 \pm 0.0003~{\rm ps^{-1}K^{-2}}$~(b).
These scattering rates are identical to within
experimental errors, even though the
AMROs in (a) and (b)
are produced by different FS sections.}
\label{fig4}
\vspace{-5mm}
\end{figure}

Using these $\omega$s,
the $\Delta R_{zz}/<R_{zz}>$ values of 
{\it all} of the AMRO
features at a particular $\phi$ 
collapse onto a single curve as a function of
$\omega \tau$; an example is shown in Fig.~\ref{fig3}(d)
for Yamaji oscillations due to the Q2D FS.
Once this correspondence is known for each $\phi$, 
$\tau$ can be readily found from AMRO data.
The $\tau$ values thus obtained follow a
$T$-dependence of the form $\tau^{-1}=\tau_0^{-1}+A T^2$ rather
closely. Moreover, to within experimental errors, 
the scattering
rates have the same $\tau_0$ and values of $A$, 
irrespective of the FS orbits involved.
This is illustrated in Fig.~\ref{fig4},
which shows $\tau$ deduced from AMROs at
$\phi=160^{\circ}$ ((a)- due to Q1D FS sections) and
$\phi = 90^{\circ}$ ((b)- due to Q2D FS sections).
In spite of the fact that the quasiparticle orbits
involved are very different, and involve distinct
FS sections, the
$T$ dependence of $\tau$ for both is virtually identical.
This suggests 
that mechanisms for superconductivity
in organic metals
that invoke a large variation
in scattering rate over the FS ({\it e.g.}
``FLEX'' methods~\cite{flex})
are inappropriate for \cuscn.
For all FS orbits studied,
the inferred $T=0$ scattering time
($\tau_0 =3.4\pm 0.2$~ps- see Fig.~\ref{fig4}) 
is close to values measured by other
means~\cite{condprob}. 

The $T^2$ dependence of $\tau^{-1}$ 
suggests electron-electron
scattering~\cite{ashcroft}. 
A $T^2$ scattering rate was
inferred from $B=0$ resistivity
measurements~\cite{ishiguro}. 
However, problems in deconvolving the
in-plane resistivity component $\rho_{||}$ from $\rho_{zz}$ in
experimental data~\cite{review,condprob}, and the influence of the
broad superconducting transition 
on the $T$-dependence of the measured
resistivity~\cite{review} mean that this attribution can
not be considered conclusive. 
By contrast, the $T$-dependent AMROs
provide an unambiguous gauge of the scattering rate of
normal-state quasiparticles, allowing the  
mechanism to be definitively identified.

When $k_{\rm B}T>4t_{\bf a}$, the interlayer 
contribution to the bandwidth, 
the FS shown in Fig.~\ref{fig2}(c) 
is ``blurred'' on a wavevector scale 
$k_{\rm B}T/\hbar v_{\rm F}$
(where $v_{\rm F}$ is the Fermi velocity)
exceeding the amplitude of interlayer corrugation.
Nevertheless, a Fermi-liquid
picture predicts that the semiclassical dynamics 
of each electron are still
governed by an equation of motion which 
contains and reflects the detailed
nature of the interlayer dispersion. 
Quasiparticle states within $\sim k_{\rm B}T$ of
the chemical potential continue to contribute
to the AMROs and SQUIT, as long as 
the underlying dispersion
is not affected by a variation in electron energy on 
the scale of $k_{\rm B}T$;
this will hold when 
$k_{\rm B}T\ll t_{||}\sim 10-100$~meV~\cite{goddardprb2004}.
Then, the only effect of raising $T$ is from
$T$-dependent contributions to the scattering rate. 
This is exactly
what we find in our AMRO data, and 
demonstrates that a coherent 3D FS
picture is valid for this compound, 
at least at $T\ltsim 0.1 t_{||}$
(a limit our experiments do not exceed), 
at which point a crossover to an
incoherent regime might be expected~\cite{mnm}.

In summary, the orientation dependence
of the interlayer 
magnetoresistance $\rho_{zz}$
of \cuscn ~has been studied in
fields of up to 45~T as a function of temperature $T$.
The ``SQUIT'' or coherence peak in $\rho_{zz}$
seen in exactly in-plane fields, a definitive
signature of a 3D Fermi surface ({\it i.e.} interlayer
coherence), persists to temperatures exceeding the 
proposed Anderson
criterion for incoherent transport 
(Eq.~1~\cite{anderson}) by a
factor $\sim 30$. Features in
the magnetoresistance have been successfully
modelled using an approach based on
field-induced quasiparticle paths on
a 3D Fermi surface, to yield the
temperature dependence of the quasiparticle
scattering rate.
The results suggest that the average scattering
rate does not vary strongly over the Fermi surface,
and that it follows a $T^2$ dependence attributable to
electron-electron scattering.

This work is funded by US Department of
Energy (DoE) LDRD grants 20040326ER
and 20030084DR and by EPSRC (UK).
NHMFL is supported by the
National Science Foundation, DoE and the
State of Florida, and HFML
Nijmegen by EuroMagNET (EU
contract RII3-CT-2004-506239).
Work at Argonne was
supported by the Office of Basic Energy Sciences,
US DoE (contract W-31-109-ENG-38).
PAG and AA acknowledge support from
the Glasstone Foundation and the Royal
Society respectively. 
Neil Harrison, Ross McKenzie and 
Nic Shannon are thanked for
helpful discussions.


\begin{thebibliography}{99}
\bibitem{vestige}
In fact the data show that
some vestiges of a 3D FS
persist up to at least 50~K~\cite{us}.
\end{thebibliography}
\end{document}